\documentstyle[aps,multicol,epsfig]{revtex}       

\begin{document}  
\begin{multicols}{2}   
\narrowtext

\noindent
{\bf The nature of attraction between like charged rods}

The origin of attractive interactions
in polyelectrolyte solutions has provoked a lot of attention.  
In recent Letters Ha and Liu \cite{HaLiu} attempted to study this interesting
problem using a two state model originally introduced by Oosawa and
Manning.  In the context of this model the counterions are treated
as either free or condensed.  

To calculate the free energy of interaction between two rod-like polyions
the authors performed the Hubbard-Stratonovich transformation to map
the Hamiltonian of the interacting charged rods onto an effective
field theory. The authors proceeded to linearize the resulting action,
finishing up with the so
called Gaussian approximation. The attraction between two charged rods, within
this approximation,  arises
as a result of correlations in {\it thermal} fluctuations.
Although nowhere stated explicitly,
the Gaussian
approximation, being a form of high temperature expansion (weak coupling),
is {\it only} valid for  $\alpha^2 \xi  \ll 1$, where $\xi$
is the Manning parameter, $\xi=\beta q^2/D b$, and
$\alpha$ is the valence of the counterions \cite{arenzon}.
However, in order to have a significant counterion condensation
it is necessary to have   $ \xi > 1/ \alpha $.  Thus, the inequality above
is strongly violated for multivalent counterions.  We must,
therefore, conclude that the Gaussian approximation is inapplicable
to the problem of attraction between like charged polyions. 
An analogy with a two component plasma $(TCP)$ is perhaps worthwhile.
It is well known that the  $TCP$ is isomorphic to the
sine-Gordon field theory.  If a Gaussian approximation is applied  
to study this field theory, a lot of important physics associated with
the strong coupling limit would be
lost.  In particular, no account could be
taken of ionic association which becomes prevalent as soon as
the electrostatic energy of 
interaction  becomes comparable to the thermal energy.  Thus, although
the Gaussian approximation is sufficient to capture the physics leading
to Debye-H\"uckel limiting laws, it is insufficient to go beyond this
point.
 
\begin{figure}[h]
\centerline{\epsfig{file=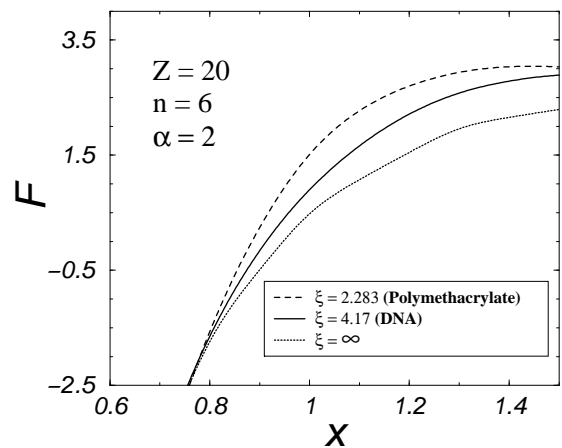,width=7.2cm} }
\caption{Force between polyions as a function of distance (in units of
$b$) for various temperatures, attraction when $F<0$.}
\label{figura}
\end{figure}

To further explore the extent of validity of the
Gaussian approximation as it applies to rigid polyions, we study a 
simple model of interaction between two charged rods with
$Z$ negative sites (separated by $b$)
and $n$ condensed multivalent counterions~\cite{arenzon}.
The Hamiltonian for this system takes a particularly simple form which can
be studied exactly. We find that the attraction is produced by
the structure of the ground state and {\it not} by the 
thermal fluctuations.
In fact for values
of Manning parameter relevant for real polyelectrolytes such as DNA,
the attractive force between the two rods is almost exactly the same
as at zero temperature, $\xi=\infty$ (Fig.\ref{figura}). The mechanism of 
attraction at zero temperature
is quite straightforward. To minimize the {\it electrostatic} energy,
the condensed counterions on the two rods arrange themselves
in a staggered configuration so that if the site of one polyion is
occupied by a counterion, a parallel site of the second polyion  stays
empty.  At short distances, this results in attraction if the number
of condensed counterions is sufficiently large, $ 2n\alpha > Z$, i.e. the
number of favorable counterion-monomer interactions is larger than the
number of unfavorable bare monomer-monomer interactions.  We find that
the same mechanism persists to finite temperature and is not 
captured in the high temperature Gaussian 
approximation as  applied by Ha and Liu.
Finally, we would like to stress that the formation of an 
ionic crystal is not
necessary for the appearance of attraction, the correlations
in positions of condensed counterions on the two rods
are sufficient.  As an example let us consider two
rods of length $Z (even)$ with $n=Z/2$ divalent condensed counterions.
Clearly the ground state configuration is $(...+-+-+...)$ on one
polyion and $(...-+-+-...)$ on the other, i.e. the counterions at zero 
temperature
form a crystal.  Obviously the  Hamiltonian
in this example is invariant under the transformation $+ \rightleftharpoons -$.
For a  finite system this symmetry cannot be broken, so that for any
 non-zero
temperature the average charge on any site is zero and no crystal is
possible.  The
correlations, however, persist and result in attraction. 
The attractive force is purely correlational, 
and to a  good approximation arises from the  
structure of the ground state.
The thermal fluctuations serve only to {\it diminish} this  attraction.

\noindent
Yan Levin$^1$, Jeferson J. Arenzon$^1$ and J\H{u}rgen F. Stilck$^2$ \\
$^1$Instituto de F{\'\i}sica, UFRGS\\ CP 15051, 91501-970, Porto Alegre, RS,
Brazil  \\
$^2$Instituto de F{\'\i}sica, UFF  \\ Av. Litor{\^a}nea, 
s/n$^o$, 24210-340, Niter{\'o}i, RJ,
Brazil

\end{multicols}     

\end{document}